\documentclass[11pt]{article}
\usepackage{epsfig}

\newcommand{\BABARPubYear}    {04}

\newcommand{\BABARProcNumber} {18}
\newcommand{\SLACPubNumber} {10783}

\input babarsym

\long\def\inst#1{\par\nobreak\kern 4pt\nobreak
    {\it #1}\par\vskip 10pt plus 3pt minus 3pt}

\begin{document}
{\pagestyle{empty}

\begin{flushright}
SLAC-PUB-\SLACPubNumber \\
\babar-PROC-\BABARPubYear/\BABARProcNumber \\
October, 2004 \\
\end{flushright}

\par\vskip 4cm

\begin{center}
\Large \bf Search for \boldmath{\Dz--\Dzb} Mixing and Rare Charm
  Decays
\end{center}
\bigskip

\begin{center}
\large 
U. Egede\\
Imperial College London \\
London, SW7~2AZ, \\
United Kingdom \\
(for the \lbabar\ Collaboration)
\end{center}
\bigskip \bigskip

\begin{center}
  \large \bf Abstract
\end{center}
Based on a dataset acquired by the \babar\ experiment running on and near the
\Y4S resonance from 1999-2002, an upper limit is set on the rate of \Dz--\Dzb
mixing using the decay mode $\Dstarp \to \Dz\pip$, followed by a semi-leptonic
decay of the \Dz. Results are compared to previous \babar\ analysis using
hadronic decays. We also set limits on the flavor-changing neutral current
decays $\Dz \to \epem$ ($\mup\mu^{-}$) and the lepton-flavor violation decays
$\Dz \to \epm \mu^{\mp}$.
\vfill
\begin{center}
Contributed to the Proceedings of the 5$^{th}$ Rencontres du Vietnam, \\
August 2004, Hanoi, Vietnam
\end{center}

\vspace{1.0cm}
\begin{center}
{\em Stanford Linear Accelerator Center, Stanford University, 
Stanford, CA 94309} \\ \vspace{0.1cm}\hrule\vspace{0.1cm}
Work supported in part by Department of Energy contract DE-AC03-76SF00515.
\end{center}

\section{Overview}
The \babar\ experiment, which is documented in detail
elsewhere\cite{Aubert:2001tu}, has since its start in 1999 not only given
results on $B$-physics but also a series of new results in charm physics.  With
a \ccbar cross section of $1.3~\nb$ at the \Y4S resonance compared to the cross
section of around $1.1~\nb$ for $B$ production there is in fact a higher prompt
charm production than $B$ production.

Here we present two new results from \babar. The first one is a search for
mixing between the neutral $D$ meson states in the semi-leptonic decay
channel\cite{Aubert:2004bn} while the other is a search for rare lepton decays
of the neutral $D$ meson\cite{Aubert:2004bs}. Both are processes that, if seen
with the current statistics, would be clear signs of physics beyond the
Standard Model~(SM).

\section{\boldmath{\Dz--\Dzb} mixing}
Charm mixing is characterised by the two parameters $x \equiv \Delta m/\Gamma$
and $y \equiv \Delta\Gamma /2\Gamma$, where $\Delta m$ ($\Delta\Gamma$) is the
mass (width) difference between the the two neutral $D$ mass eigenstates, and
$\Gamma$ is the average width. We define the overall time-integrated mixing
rate as $R_{mix}=(x^2+y^2)/2$.

Mixing between the neutral charm mesons is, within the SM, heavily suppressed
by the GIM mechanism. The expected mixing rate through box and di-penguin
diagrams is $\order(10^{-8}-10^{-10})$ but enhancements involving
non-pertubative effects are possible. For a recent review of predictions for
both the SM rate and possible New Physics contributions
see\cite{Petrov:2003un}.

To search for mixing the production flavour of the $D$-meson is tagged from
the charge of the pion in the decay $\Dstarp \to \Dz\pip$ and the decay
flavour is tagged from the charge of the electron in the decay $\Dz \to \Km
\ep \nue$.  Charge conjugation is implied everywhere. The decay where the pion
and the electron have opposite charge (called the \emph{wrong sign} mode), can
only proceed when the \Dz oscillates into a \Dzb before its decay.  The
\emph{right sign} mode where the pion and electron have the same charge is
used as a normalisation mode. In an analysis where the efficiency for right
sign and wrong sign decays are identical, $R_{mix}$ is simply given as the
time-integrated ratio of the two decay modes.

The analysis is based on a sample of $87~\invfb$ and uses the $\Dz \to \Km \ep
\nue$ sample while ignoring the less pure muon sample. The mass difference
$\Delta M$ between the partially reconstructed \Dstarp candidate and the
partially reconstructed \Dz candidate is together with particle identification
the main selection criterion to obtain a pure sample. 

Separate neural networks, with input parameters specifically describing the
\Dz daughters and globally describing the rest of the event, are used to
select signal events and reconstruct the \Dz momentum vector. The neural
networks, combined with charged kaon and electron particle identification,
provide a relatively pure selection of unmixed signal events and give a
resolution in $\Delta M$ of $2.2~\mevcc$.

The time distribution of the right sign control sample follows a simple
exponential convoluted with a resolution function $R$, while the wrong sign
signal has the form
\begin{equation}
  \label{eq:WSsignal}
  \Gamma_{WS}(t) = e^{-t}\frac{R_{mix}}{2}t^2 \otimes R\ ,
\end{equation}
where $t$ is measured in units of the \Dz lifetime.

\begin{figure}[htb]
  \begin{minipage}[t]{0.48\linewidth}
    \psfig{figure=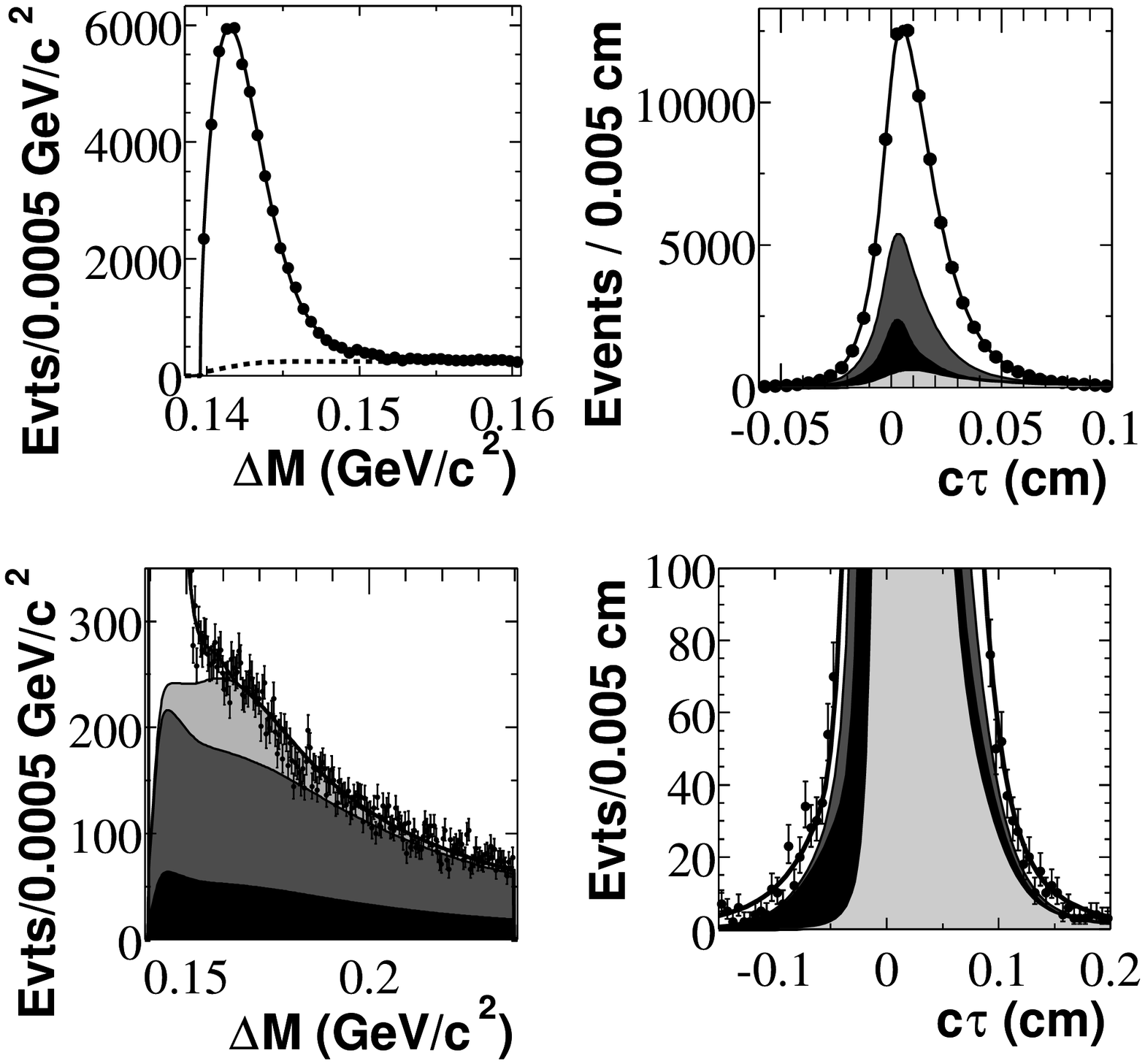,width=\linewidth}
    \caption{\label{fig:rsfit}$\Delta M$ (left) and decay time (right)
      projections of fit (solid lines) to RS data (points): (top left) $\Delta
      M$ signal region --- unmixed signal (above dashed line), background
      (dashed line); (bottom left) magnified vertical scale $\Delta M$ full
      fit region --- unmixed signal (white), $D^{+}$ background (light grey),
      $D^{0}$ background (dark grey), zero-lifetime background (black); (top
      right) decay time signal region --- signal and background components as
      in bottom left plot; (bottom right) magnified vertical scale decay time
      full fit region --- signal and background components as in bottom left
      plot of this figure.}
  \end{minipage}
  \hfill
  \begin{minipage}[t]{0.48\linewidth}
    \psfig{figure=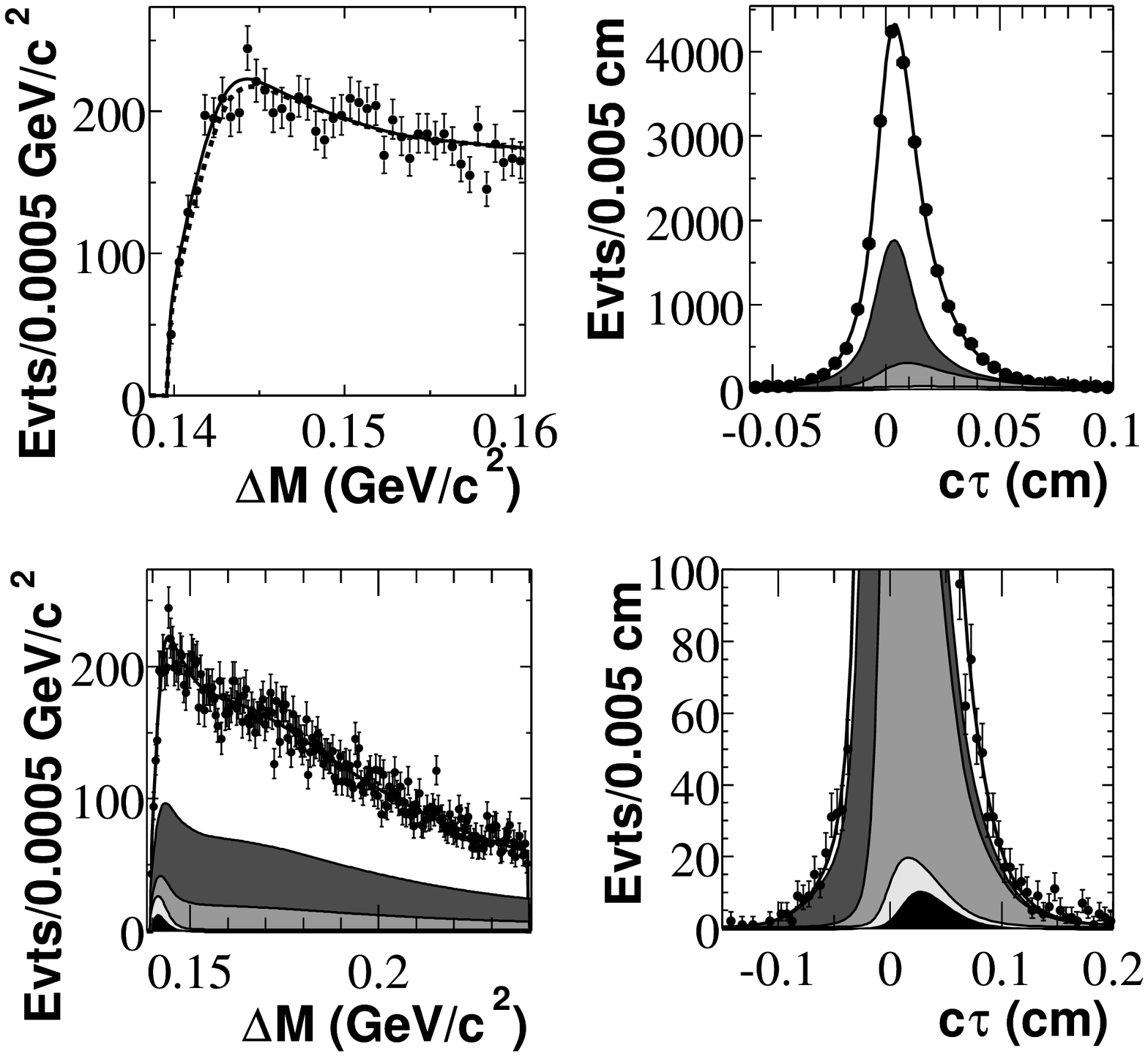,width=\linewidth}
    \caption{\label{fig:wsfit}$\Delta M$ (left) and decay time (right)
      projections of fit (solid lines) to WS data (points): (top left) $\Delta
      M$ signal region --- mixed signal (above dashed line), background
      (dashed line); (bottom left) $\Delta M$ full fit region --- $D^{0}$
      background (white), zero-lifetime background (dark grey), non-peaking
      $D^{+}$ background (intermediate grey), peaking $D^{+}$ background
      (light grey), mixed signal (black); (top right) decay time signal region
      --- signal and background components as in bottom left plot; (bottom
      right) magnified vertical scale decay time full fit region --- signal
      and background components as in bottom left plot.}
  \end{minipage}
\end{figure}
An unbinned extended likelihood fit is performed on the 2-dimensional
distribution of signal and background in the variables $t$ and $\Delta M$;
first on the right sign sample and then, with the shared parameters between
the two datasets constrained, on the wrong sign sample. In
Figs.~\ref{fig:rsfit} and~\ref{fig:wsfit}, projections of the fit can be seen
overlaid on the data. The result is a wrong sign signal yield of $114 \pm 61$
events.

Systematic errors arise mainly from the assumptions related to the shape of
signal and background in the $\Delta M$ variable and when added in quadrature
add up to 34\% of the statistical error. Combining the wrong sign yield with
the right sign yield we get the final result
\begin{eqnarray}
  \label{eq:mixResult}
  R_{mix} & = & 0.0023 \pm 0.0012\stat \pm 0.0004\syst \\
  R_{mix} & < & 0.0042\mbox{~at 90\% CL.}
\end{eqnarray}
Systematics are taken into account by scaling the log likelihood curve for the
fit to the wrong sign yield with the systematic error added in quadrature
($\sqrt{1+0.34^2} = 1.06$). The upper limit was calculated assuming a flat
prior for the number of wrong sign events to be positive. In
Fig.~\ref{fig:mixingCompare}, the result is compared to previous results.

\section{Flavour-changing neutral current and lepton-flavour violating decays}
In this analysis, a search is performed for the flavour-changing neutral
current~(FCNC) decays $\Dz \to \epem$ and $\Dz \to \mup\mu^{-}$ and the
lepton-flavour violating~(LFV) decays $\Dz \to \epm \mu^\mp$. In the SM, the
FCNC decays are highly suppressed by the GIM mechanism and the LFV decays are
strictly forbidden. Compared to rare decay searches in the $K$ and $B$ sector,
rare $D$ decays are sensitive to new physics involving the up-quark sector
such as certain $R$-parity violating supersymmetric
models\cite{Burdman:2001tf}.

As in the previous analysis the \Dz is required to originate from a \Dstarp,
but this time to ensure as clean a sample as possible. For the same reason, the
\Dz is required to have a momentum above $2.4~\gevc$ in the \Y4S centre-of-mass
frame to reduce background from combinatorics involving the decay products of
$B$ mesons. Electrons~(muons) are identified with an efficiency of 95\%~(60\%)
with a hadron misidentification probability of 0.2\%~(2\%) as measured on a
$\tau$ decay control sample.

The decay $\Dz \to \pip\pim$ is used as a control sample as it has very
similar kinematics and, as such, the systematic errors can be minimised. Apart
from the particle identification, the selection of the control channel, is
identical to the criteria used for the signal.

\begin{figure}
  \begin{minipage}[t]{0.48\linewidth}
    \psfig{figure=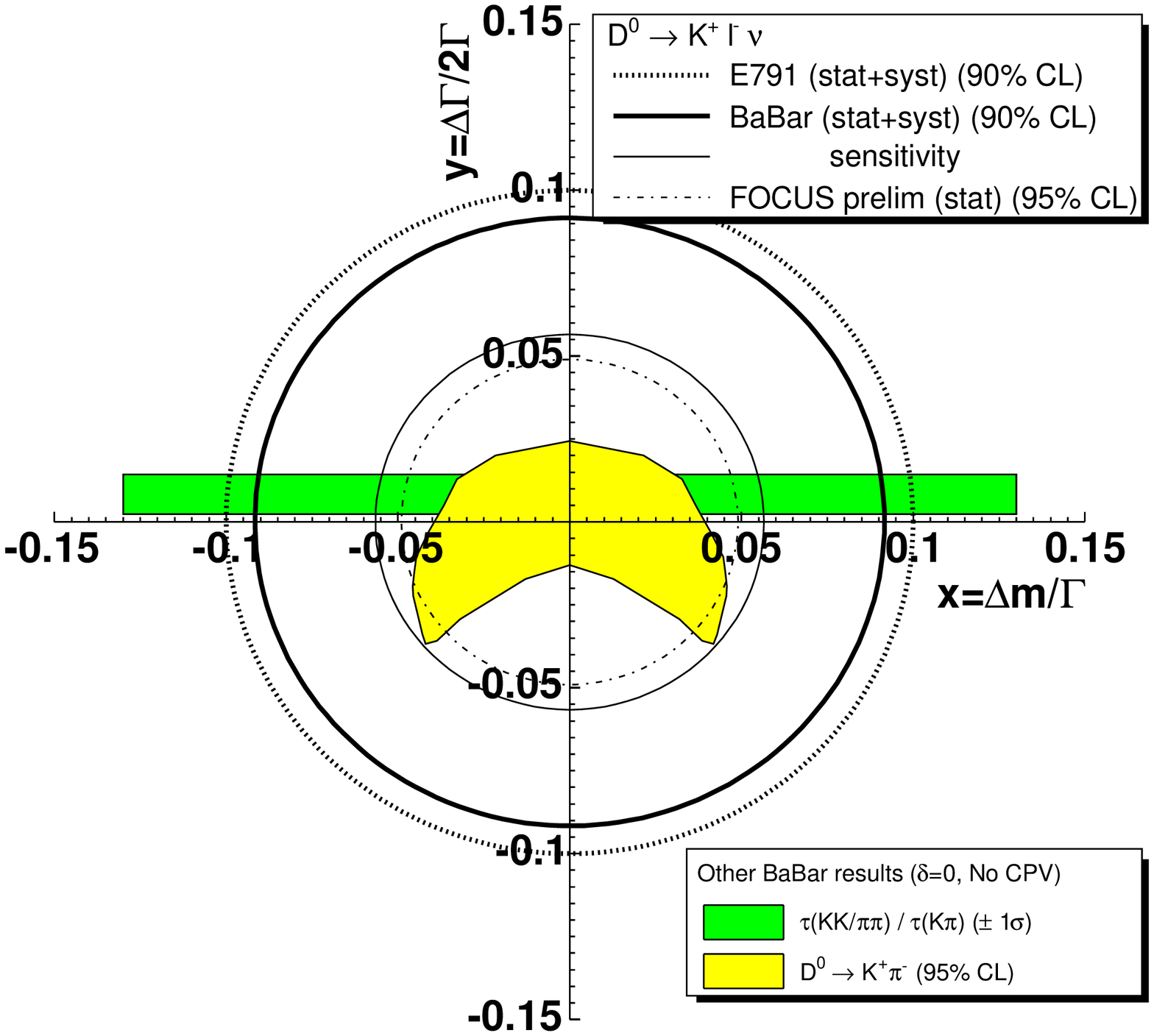,width=\linewidth}
    \caption{The \babar\ limit compared to other preliminary and published
      results using the semi-leptonic decay to search for
      mixing\protect\cite{Aitala:1996vz,ICHEP:2002}.  The thin solid circle
      shows the sensitivity of the \babar\ analysis if zero wrong sign signal
      events were seen. For the comparison to the \babar\ results from a
      mixing search with hadronic decays
      modes\protect\cite{Aubert:2003pz,Aubert:2003ae} it is assumed that
      there is no \CP\ violation and that the strong phase difference is zero.
      \label{fig:mixingCompare}}
  \end{minipage}
  \hfill
  \begin{minipage}[t]{0.48\linewidth}
    \psfig{figure=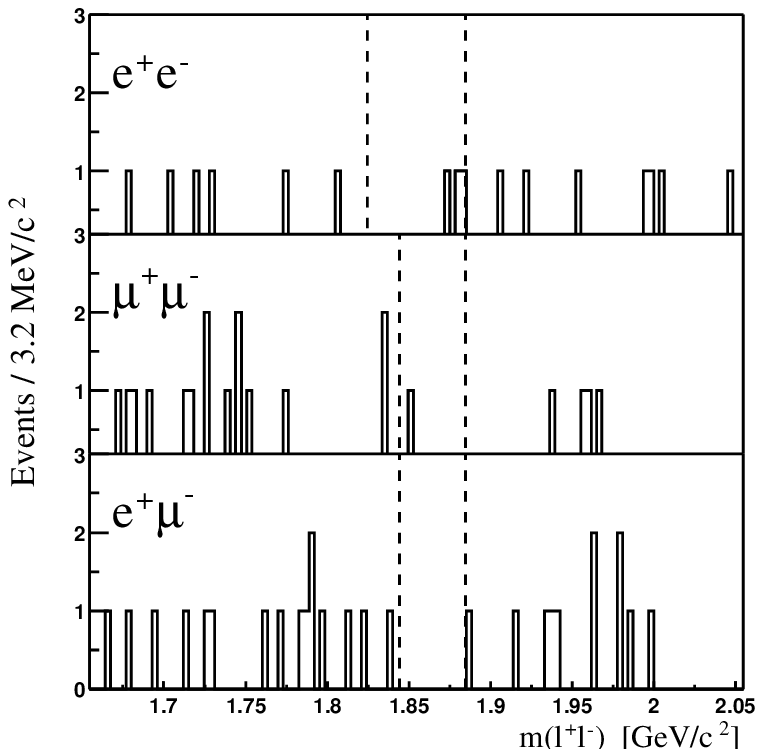,width=\linewidth}
    \caption{The dilepton invariant mass distribution for each of the decay
      modes. The dashed lines indicate the optimised signal mass windows.
      \label{fig:rare}}
  \end{minipage}
\end{figure}
Based on a sample of 122~\invfb and after optimisation of the selection
criteria the events seen in Fig.~\ref{fig:rare} remain. The background is
estimated from the sidebands with a looser selection applied and then scaling
it to the final selection taking the small correlation between the criteria
into account.

We do not see any signal in any of the channels and the branching fraction
upper limits have been calculated using an extension of the Feldman-Cousins
method\cite{Conrad:2002kn} that avoids the unwanted effect of the
Feldman-Cousins method\cite{Feldman:1997qc} that the UL for a search can go
down in case of an upwards fluctuation in the expected background. Our result
and a comparison to previous published results\cite{Aitala:1999db,Abt:2004hn}
can be seen in Table~\ref{tbl:br}.

\begin{table}[htb]
\begin{center}
\begin{tabular}{lccc}\hline\hline
 & $\Dz\to e^+e^-$ & $\Dz\to \mu^+\mu^-$ & $\Dz\to e^\pm\mu^\mp$
\\\hline
$N^{hh}_{\rm bg}$ & $0.02$ & $3.34\pm0.31$ & $0.21$ \\
$N^{\rm comb}_{\rm bg}$ & $2.21\pm 0.38$ & $1.28\pm0.32$ & $1.93\pm0.36$ \\
$N_{\rm bg}$ & $2.23\pm 0.38$ & $4.63\pm0.45$ & $2.14\pm0.36$ \\
$N_{\rm obs}$ & 3 & 1 & 0 \\
UL at 90\% CL & $1.2\times10^{-6}$ & $1.3\times10^{-6}$
& $8.1\times 10^{-7}$ \\\hline
Previous published limit\protect\cite{Aitala:1999db,Abt:2004hn} 
 & $6.2\times 10^{-6}$ &  $2.0\times 10^{-6}$  & $8.1\times 10^{-6}$ 
\\\hline\hline
\end{tabular}
\caption{The summary of the number of expected background
  events ($N_{\rm bg}$), number of observed events ($N_{\rm obs}$), and the
  branching fraction upper limits at the 90\,\% confidence level for each
  decay modes.  The uncertainties quoted here are total uncertainties.  The
  uncertainty of $N^{hh}_{\rm bg}$ is negligible for the $ee$ and $e\mu$ decay
  modes.}
\label{tbl:br}
\end{center}
\end{table}

\end{document}